\documentclass[aps,prl,twocolumn,showpacs,superscriptaddress,groupedaddress]{revtex4}  
\usepackage{graphicx}  
\usepackage{dcolumn}   
\usepackage{bm}        
\usepackage{amssymb}   
\usepackage{epstopdf}
\usepackage{braket}
\begin{document}
\widetext


\title{Magnetic Trapping of Molecules via Optical Loading and Magnetic Slowing}

\author{Hsin-I Lu}
\email{lu@cua.harvard.edu} 
\affiliation{School of Engineering and Applied Sciences, Harvard
  University, Cambridge, MA 02138, USA}
\affiliation{Harvard-MIT Center for Ultracold Atoms, Cambridge,
  Massachusetts 02138, USA}
\author{Ivan Kozyryev}
\affiliation{Department of Physics, Harvard University, Cambridge,
  Massachusetts 02138, USA}
\affiliation{Harvard-MIT Center for Ultracold Atoms, Cambridge,
  Massachusetts 02138, USA}
\author{Boerge Hemmerling}
\affiliation{Department of Physics, Harvard University, Cambridge,
  Massachusetts 02138, USA}
\affiliation{Harvard-MIT Center for Ultracold Atoms, Cambridge,
  Massachusetts 02138, USA}
\author{Julia Piskorski}
\affiliation{Department of Physics, Harvard University, Cambridge,
  Massachusetts 02138, USA}
\affiliation{Harvard-MIT Center for Ultracold Atoms, Cambridge,
  Massachusetts 02138, USA}
\author{John M. Doyle}
\affiliation{Department of Physics, Harvard University, Cambridge,
  Massachusetts 02138, USA}
\affiliation{Harvard-MIT Center for Ultracold Atoms, Cambridge,
  Massachusetts 02138, USA}

\begin{abstract}
Calcium monofluoride (CaF) is magnetically slowed and trapped using optical pumping. Starting from a collisionally cooled slow beam, CaF with an initial velocity of $\sim$\,30 m/s is slowed via magnetic forces as it enters a 800 mK deep magnetic trap. Employing two-stage optical pumping, CaF is irreversibly loaded into the trap via two scattered photons. We observe a trap lifetime exceeding 500 ms, limited by background collisions. This method paves the way for cooling and magnetic trapping of chemically diverse molecules without closed cycling transitions.
\end{abstract}

\pacs{ 37.10.Mn 37.10.Pq 37.10.Vz}
\maketitle
\begin{figure*}[]
  \centering
    \includegraphics[width=6.7in]{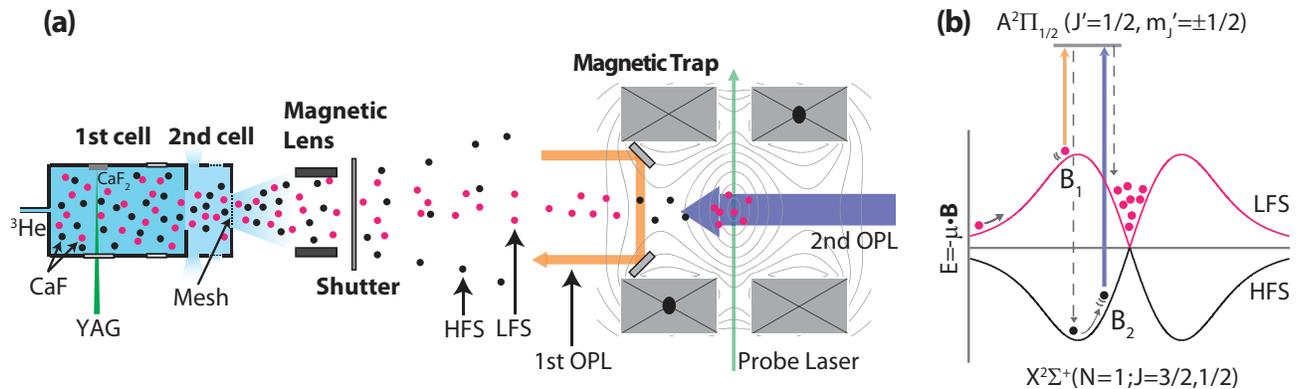}
    \caption{(a) Schematic of the apparatus (not to scale). 
    A slow CaF beam is produced from a two-stage cell. The LFS (solid pink circles) are focused by a magnetic lens at 6 cm from the cell exit aperture and then enter a superconducting magnetic trap at 30 cm downstream. The HFS (solid black circles) in the beam diverge rapidly after being defocused by the magnetic lens.
    Two optical pumping lasers (OPL) are used to achieve irreversible loading: one interacts with the LFS near the saddle point; the other with the HFS inside the trap. A cryogenic shutter in between the magnetic lens and trap blocks the buffer gas after trap loading. Loaded CaF is detected by sending a probe laser at 531 nm (via frequency doubling of a diode laser at 1062 nm) through the trap midplane. The fluorescence is collected by two fiber bundles (not shown here) mounted on the magnet cask and recorded by a photomultiplier tube. (b) Optical loading scheme for molecules with a magnetic dipole moment of 1 $\mu_B$ into a magnetic trap. Potentials experienced by the LFS and HFS are represented in pink and black curves. Here, we denote the states with the quantum numbers of CaF.}
    \label{apparatus}
\end{figure*}

Experiments using cooling and trapping methods continue to shed light on the physics of molecules, especially collisions. Cold molecular interactions, sometimes sensitive to external fields, have been studied using buffer-gas loading and atom association methods \cite{CampbellPRL2009,Hummon2011,Zirbel2008,Ni2010}.
Recently, novel collisions have been studied using trap loss techniques \cite{Sawyer2011,Hummon2011,Parazzoli2011} and used in the evaporative cooling of polar molecules \cite{Stuhl2012}.
Ultra-cold molecules produced by assembling two laser cooled atoms \cite{Kerman2004,Ni2008} enable the study of ultra-cold chemistry and electric dipole interactions in the quantum regime \cite{Ni2010,Ospelkaus2010,deMiranda2011}.
Current experiments searching for physics beyond the Standard Model also employ new cold molecule methods \cite{Vutha2010,Hudson2011}.

New insights into fundamental collision processes of molecules could be gained by extending trapping to new, more complex species. Among the various cooling methods, cold beam techniques provide the most diverse sources of cold (around 1 K), neutral molecules \cite{Lemeshko2013,Meerakker2005,Hoekstra2007,Momose2013,Narevicius2008_II,vanBuuren2009,Chervenkov2014,Hutzler2011,Barry2011}. Trapping of molecules from these sources would allow for longer interaction times and detailed study of collisions. Loading of traps has been demonstrated for Stark decelerated and electrically filtered molecular beams \cite{Bethlem2002,Englert2011,Stuhl2012,Zeppenfeld2012}. The buffer-gas cooled beam is very general, and easily produces high molecular flux for many different species, including radicals \cite{Hutzler2012}. However, irreversible trap loading from such a beam has not been realized. Scattering of photons (i.e.~optical pumping, which can irreversibly drive molecules between untrapped and trapped states) has been employed to load laser cooled cesium and chromium atomic beams into ac magnetic and hybrid traps \cite{Cornell1991,Falkenau2011}. Such a loading scheme is potentially suitable for a large class of molecules, including those without closed cycling transitions, and thus could open the way to trapping of more complex molecular species, such as polyatomic molecules. This could be particularly important for new experiments ranging from study of strongly interacting quantum systems to chemical and particle physics.


Here, we demonstrate a general trapping method for magnetic molecules by loading a collisionally cooled slow beam of radical molecules into a magnetic trap, using a two-stage optical pumping scheme. The buffer-gas cooled CaF beam used in this work has a peak forward velocity of $v_f=$~55 m/s with velocity width 45 m/s. Optical pumping in conjunction with the magnetic field leads to slowing and trapping of CaF. CaF molecules in the $X^2\Sigma^+ (v=0,N=1)$ state are observed in the trap for longer than 1\,s.
The spontaneously emitted photons carry away the potential energy and entropy of the molecules and hence the loading mechanism is dissipative and irreversible, which could allow for build up of phase space density \cite{Cornell1991,Falkenau2011}.

Buffer-gas cooling, which exploits elastic collisions of molecules with cold inert gases, offers a general approach to generate cold molecules at a temperature of $\sim$ 1 K \cite{Campbell2009}. Several magnetic species, including CaH, NH, CrH, and MnH, have been trapped via in situ loading in a buffer-gas cell inside a magnetic trap \cite{Weinstein1998,Campbell2007,Stoll2008,Tsikata2010}. The constraint on the number of collisions needed during trap loading limits the in-cell buffer-gas loading method to magnetic species with a ratio of elastic to inelastic He-molecule cross sections of $\gamma >10^4$ \cite{Hancox2004}.
By contrast, the loading step reported here occurs at a low buffer-gas density of $<10^{12}$ cm$^{-3}$, which makes it more general and applicable to low $\gamma$ molecules.

Typical buffer-gas molecular beams have a moderate $v_f$\,$\sim$\,150 m/s, emitting from a single-stage cell with hydrodynamic enhancement \cite{Hutzler2011,Barry2011}.~Direct laser cooling, slowing, and 2D magneto-optical trapping of molecules have recently been demonstrated with such single-stage buffer-gas beams \cite{Shuman2010,Barry2012,Hummon2013}. Recently, a low velocity CaH beam with $v_f$\,$\sim$\,60 m/s was reported, based on an advanced buffer-gas cell design (two-stage cell) \cite{Lu2011}. This offered the possibility for direct trap loading of slow molecules. A similar two-stage cell is employed here to generate our slow CaF beam.

A key challenge in all trap loading experiments is the rapid divergence in the beam as it is slowed, leading to too few molecules or atoms in the trapping region. In this work, we directly solve this problem by having the final stage of slowing in the trapping region. Because only a few photon scattering events are sufficient for trap loading, this method has a significant advantage in trapping magnetic molecules with non-diagonal Franck-Condon factors, for which direct laser cooling is poorly suited.

The apparatus is depicted in Fig.~\ref{apparatus}(a).~CaF is created by laser ablation of a CaF$_2$ solid precursor inside the two-stage cell~\cite{Lu2011} at 1.3\,K. CaF molecules thermalize with cold $^3$He at a density of $n_{\text{1,He}}\sim$\,$10^{15}$ cm$^{-3}$ in the 1st cell and then enter the 2nd cell, which has $n_{\text{2,He}} \sim n_{\text{1,He}}/10$ and a piece of mesh on its exit aperture. We produced cold beams successfully with both $^4$He and $^3$He; the $^3$He based source gave about a factor of two higher flux and about 30 m/s lower velocity than the $^4$He source.
Based on the systematic studies reported in Ref.~\cite{Lu2011}, the value of $n_{\text{2,He}}$ used here provides just enough collisions for slowing with a modest reduction in molecular flux compared to the single-stage source. (The effect of the mesh is described in detail elsewhere \cite{Lu2011}.)
A typical CaF beam has an intensity of $3\times 10^9$ molecules/sr/pulse in a duration of 10 ms for both $N=0$ and $N=1$ states.

Low-field seekers (LFS) are collimated by a hexapole magnetic lens (maximum field strength of 1\,T)
and then propagate to a quadrupole magnetic trap, which is operated at 3.5 T for loading CaF ($N=1$). A cryogenic shutter after the magnetic lens can be closed within $10$ ms, after the molecular beam passes through. When entering the trap, the LFS lose their kinetic energy while climbing up the potential hill, as illustrated in Fig.~\ref{apparatus}(b). The first optical pumping laser (OPL), resonant with the LFS near the saddle ($B_1$), optically pumps CaF to high field seekers (HFS) via the $X^2\Sigma^+(v=0)$ $\rightarrow$ $A^2\Pi_{1/2}(v'=0)$ transition at 606 nm. The HFS proceed to the trap center, get further decelerated, and are pumped by the 2nd OPL to the trappable state (LFS) at $B_2$. In principle, scattering two photons is sufficient for trap loading.
The two OPLs originate from cw dye lasers (tunable, single-frequency with a linewidth of $\sim2$ MHz) and each has a power of about 50 mW.

A necessary step for trap loading is to demonstrate the state transfer during the pumping process, which is established for CaF ($N=0$) experimentally at a field of 2.17 T. We start by monitoring the fluorescence signal of the CaF ($N=0$) beam via sending a probe laser ($X^2\Sigma^+(v=0)$ $\rightarrow$ $B^2\Sigma^+ (v'=0)$ transition at 531 nm) through the trap midplane. The HFS and LFS can be spectroscopically resolved in the field.
When the 1st OPL resonant with the LFS at the saddle is turned on, we observe a depletion of the LFS and a transfer to the HFS populations of the CaH beam at the trap center.
With the addition of the 2nd OPL, which pumps the populated HFS immediately at the saddle point, we replenish the LFS population with an efficiency of 15$\%$ for transferring LFS $\rightarrow$ HFS $\rightarrow$ LFS (see supplemental material \cite{SM2013}). For $N=0$, the efficiency of transferring between the LFS and HFS is limited by the leakage to the dark rotational excited state $N=2$.

\begin{figure}[h!]
  \centering
    \includegraphics[width=3.2in]{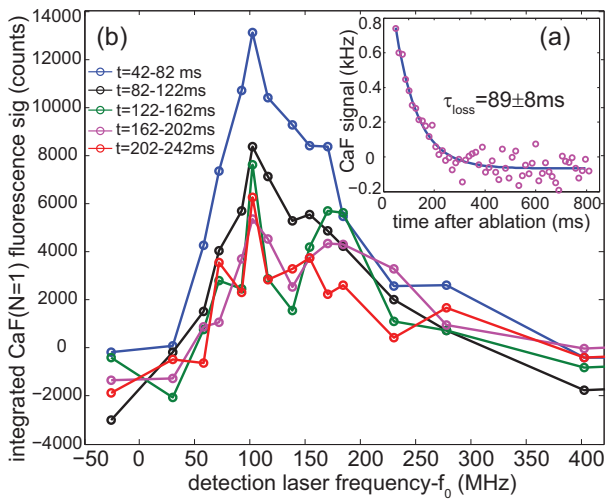}
    \caption{(a) Time decay of the trapped CaF ($N=1$) signal at the resonant frequency, relative to a background count rate taken before ablation. The buffer-gas flow is 0.75 sccm. A single exponential fit gives a decay constant of 89 $\pm$ 8 ms. (b) Spectrum of trapped CaF ($N=1$) for different time intervals relative to ablation with a He flow of 1.25 sccm. $f_0$ is the resonance frequency measured at zero field. }
    \label{Spectrum}
\end{figure}

We decide to perform the trapping experiment on $N=1$ for two reasons. First, the rotational leakage channel can be suppressed by driving a $X^2\Sigma^+(v=0,N=1)$ $\rightarrow$ $A^2\Pi_{1/2}(v'=0,J'=1/2)$ transition, as pointed out by Ref.~\cite{Shuman2009}. This leads to a more efficient state transfer than $N=0$ during the optical pumping process.
The Zeeman level of the LFS in $N=0$ ($E(N=0)+\mu_B B$) crosses that of the HFS in $N=2$ ($E(N=2)-\mu_B B$) at a field of $B_{\text{avoided}}=(E(N=2)-E(N=0))/2\mu_B=$ 2.2 T (see supplemental material \cite{SM2013}), where $E(N)$ is the rotational energy.
Due to the anisotropic hyperfine interaction, the level crossing is a true avoided crossing.
When the CaF beam passes through $B_{\text{avoided}}$ on the way to the saddle, the LFS adiabatically turn into the HFS.
On the other hand, the LFS of $N=1$ have an increased $B_{\text{avoided}}$ = 3.67 T due to a larger rotational energy difference between $N=1$ and $N=3$, allowing us to operate the trap at a higher depth than $N=0$.

The capture energy of this loading scheme can be understood as follows. Only molecules with enough kinetic energy to climb up two potential hills can reach the trap center, setting the lower bound of the capture energy to be $E_L=\mu_B \times (2B_1-B_2)$. After deceleration, molecules with kinetic energy lower than the trap depth, $E_D \sim \mu_B \times (B_1-B_2)$, can remain trapped. The capture energy is hence $E_L < E_f < E_L+E_D$, where $E_f$ is the kinetic energy of the molecules in the beam.
To load CaF ($N=1$) into the trap, we choose ($B_1$, $B_2$)=(3.5, 2.27) T, yielding $v_c=$ $29.8 - 33.5$ m/s and a trap depth of $E_D\sim$ 826 mK. We note that $B_2=2.27$ T is chosen to prevent accidentally pumping the LFS to $A^2\Pi_{1/2} (v'=0, J'=3/2)$ state by the 2nd OPL, which spatially overlaps with the CaF beam in the current setup (see also supplemental material \cite{SM2013} for the Zeeman levels of CaF in $A^2\Pi_{1/2}$ ($v'=0,J'=1/2,3/2$) states).

\begin{figure}[h!]
  \centering
    \includegraphics[width=3.05in]{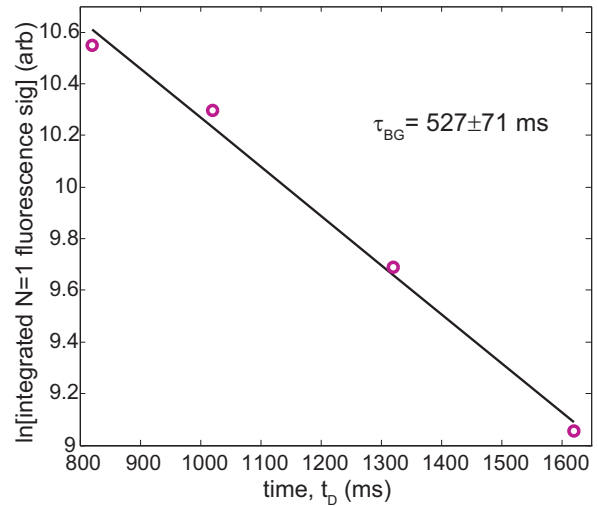}
    \caption{Integrated fluorescence signal, proportional to the number of trapped molecules, as a function of time, $t_D$. The loss is dominated by elastic collisions with cold background helium in the trap region.}
    \label{WithShutterCollisionDecay}
\end{figure}

Fig.~\ref{Spectrum} shows a main result of this Letter, which contains a time decay trace of trapped CaF ($N=1$) at the resonant frequency (Fig.~\ref{Spectrum}(a)) and a spectrum of the trapped CaF integrated over different time windows (Fig.~\ref{Spectrum}(b)). The cryogenic shutter is not used for this data set. A decay time of $\tau_{loss}=$ 89 ms in the trap is set by two factors. First, the continuously flowing $^3$He gas (0.75 sccm) from the beam source can knock out the trapped molecules, as the trap depth is comparable to the temperature of the $^3$He gas. The second effect is the result of probing trapped CaF via the $X^2\Sigma^+(v=0,N=1)$ $\rightarrow$ $B^2\Sigma^+(v'=0,N'=0)$ transition. After a few scattering events, the molecules can decay to other hyperfine states within the LFS manifold. These states stay trapped but remain dark to the probe laser.


The main limitation for achieving longer trap lifetimes is the net buffer-gas density in the trap. Reducing the background gas collisions, including decreasing the buffer-gas flow and blocking the buffer-gas beam after the molecules have traversed the loading zone, increases the lifetime.
We operate the molecular beam at a buffer-gas flow of 0.5 sccm and close the shutter after the optical loading process. To remove the artifact of optical pumping of the trapped molecules into dark states, we switch on the detection laser at different delay times, $t_D$ (see supplemental material \cite{SM2013}). By integrating the time decay signal over a duration of 200 ms, we obtain an integrated signal described by $S(t_D)\propto e^{-t_D/\tau_{BG}}$, where $\tau_{BG}$ is the decay time due to background gas collisions. In Fig.~\ref{WithShutterCollisionDecay}, we plot the integrated signal on a logarithmic scale versus $t_D$.
A fitted decay time constant of $\tau_{BG}=$ 527\emph{$\pm71$} ms is purely limited by collisions with the $^3$He gas at 4 K inside the trap.

Calculating the background He density is challenging due to the unknown pumping speed of the cryogenic pumps, the potential gas load from the imperfect blocking of the shutter, and the unknown surface desorption of helium. However, we can estimate this density using Monte-Carlo trajectory simulations that include the effect of CaF-$^3$He collisions.
When trapped CaF collides with $^3$He, three collisional processes can occur: elastic collision, spin depolarization, and rotational state changing.
These collisional properties of CaF with $^3$He at 2 K were studied by Maussang $\textit{et al.}$ \cite{Maussang2005}. It was argued that the spin depolarization of $N=1$ is much more efficient than $N=0$ and a lower bound of $\gamma_{\text{N=1}}$ $>$ $8000$ could be placed.
Including these processes in our simulation, we find a density of $n_{\text{He}}=5\times 10^{10}$ cm$^{-3}$ gives a trap lifetime of 520 ms, under the assumption of $\gamma_{\text{N=1}}=10^4$.
To distinguish the relative contribution of elastic collision and spin-depolarization, we can assume $\gamma_{\text{N=1}}=100$ in the simulation, resulting in a marginally reduced (by 20$\%$) simulated lifetime. This indicates that elastic collisions limit the observed trap lifetime.
The spatial distribution of the initially loaded molecules can also be obtained from the trajectory simulations. In combination with the light collection efficiency (determined by modeling), we estimate the number of the trapped molecules to be $\sim 2\times10^4$, corresponding to a density of $\sim10^3$ cm$^{-3}$ with an average energy $\leq E_D$ of around 800 mK.

In conclusion, we demonstrated a general trap loading scheme for magnetic molecules by employing magnetic slowing and two-stage optical pumping on a slow molecular beam. Magnetically trapped CaF ($N=1$) with a lifetime of 530 ms was realized by incorporating a cryogenic shutter for reducing the buffer-gas density in the trap. The attained trap lifetime is limited by elastic collisions with the background $^3$He gas at a density of $\sim$ 5$\times 10^{10}$ cm$^{-3}$, as indicated by the trajectory simulation.~Potential ways to improve the trap vacuum include implementing a better differential pumping design and increasing the area of cryogenic pumps inside the trap, which could increase the trap lifetime significantly.

The method we developed here opens up several possible research directions, including the study of cold collisions with more complex species.
Co-loading atomic species with molecules appears straightforward, providing a platform for studying cold atom-molecule collisions. For example, the study of trapped Li-molecule collisions is important for exploring the possibility of the proposed sympathetic cooling of molecules. A suitable molecular species for co-trapping with Li is CaH, which we have already produced in a slow beam \cite{Lu2011}, and which is predicted to have favorable collisional properties with Li in a magnetic trap with $\gamma_{Li-CaH,theory}>100$ between 10 $\mu$K $-$ 10 mK \cite{Tscherbul2011}.
Magnetically co-trapping Li and CaH also allows for demonstrating cold controlled chemistry via polarizing the electronic spins \cite{Singh2012,Tscherbul2011}.
In addition, CaH has a larger rotational constant than CaF, enabling the trap to operate at a higher depth. This also leads to a higher capture velocity that matches the peak distribution of our slow beam.
We expect to load a large number of Li atoms (compared to CaF) for several reasons, including a large ablation yield reported \cite{Singh2012} and a higher loading efficiency due to a light mass, no avoided crossings, and no leakage to dark states during the pumping process.

This method is also readily extended to other magnetic molecules, including polyatomic molecules with more vibrational degrees of freedom than diatomic molecules, which makes laser cooling infeasible. For example, CaOH and SrOH molecules--with a linear geometry in the electronic ground state ($^2\Sigma^+$), rotational constants similar to CaF, and visible transitions for optical pumping and detection--could be accumulated in a magnetic trap using the current scheme, with the scattering of only a few photons.
Rigorous theoretical calculations indicate several polyatomic species have similarly small spin depolarization rates to CaH when colliding with He \cite{Tscherbul2011sympathetic}.
Realizing magnetic trapping of polyatomic molecules is the starting point for experimentally studying the spin depolarization of polyatomic molecules in collisions with either He or $^2S_{1/2}$ (e.g. Li) atoms.
We also note that starting with two-stage buffer-gas beams in combination with a few slowing stages (such as Zeeman deceleration), the optical loading method can be applied to molecules with small rotational constants.

We thank D. Patterson for helpful discussions. This work was supported by NSF.

\section{Supplementary Material}
\subsection{State manipulation during optical pumping process}
\begin{figure}[h!]
  \centering
    \includegraphics[width=3.2in]{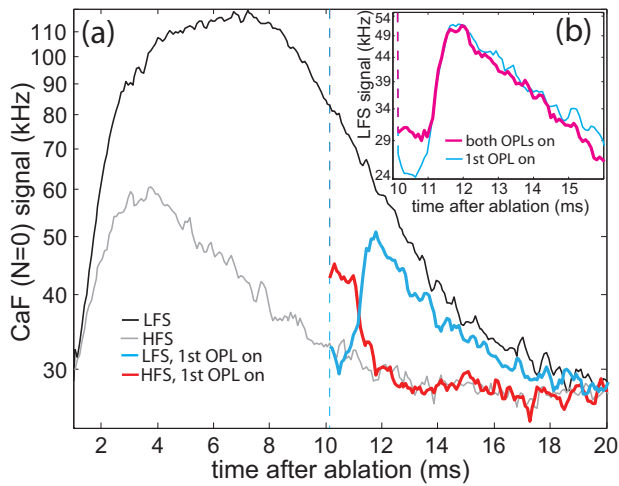}
    \caption{(a) The LFS (HFS) of CaF ($N=0$) detected at the trap midplane are shown in black (gray) without the 1st OPL. When the 1st OPL is turned on between $0-10$ ms, the PMT is overwhelmed with scattered photons. Blue and red traces represent the LFS and HFS signals when the 1st OPL is switched off (vertical dash line), demonstrating the state transfer from the LFS to HFS. (b) Switching the 2nd OPL on ($9-10$ ms) in addition to the 1st OPL ($0-10$ ms) pumps the HFS back and recovers the LFS signal between $10-11$ ms (pink trace). As a comparison, the blue trace depicts the LFS with only the 1st OPL on.}
    \label{StatePreparation}
\end{figure}

The state transfer of CaF ($N=0$) during the 1st optical pumping stage with the trap operated at 2.17 T is shown in Fig.~\ref{StatePreparation}(a). CaF ($N=0$) is detected at the trap midplane via laser induced fluorescence excited on the $X^2\Sigma^+(v=0,N=0)$ $\rightarrow$ $B^2\Sigma^+ (v'=0,N'=1)$ transition at 531 nm. We are able to spectroscopically resolve the LFS and HFS in the magnetic field, which are plotted as black and gray traces in Fig.~\ref{StatePreparation}(a). The magnetic lens focuses the LFS while defocusing the HFS, resulting in different signal heights. A typical LFS beam (without OPLs applied) arriving at the trap midplane has $10^8$ molecules/pulse. When the 1st OPL is applied between 0 and 10 ms, a dip between $10-11$ ms shown as blue trace in Fig.~\ref{StatePreparation}(a) represents the depletion of the LFS. After the pumping laser is turned off at the saddle, slow molecules take $1-2$ ms to reach the trap midplane, causing the LFS signal to recover after 11 ms.
Additional HFS molecules (red trace in Fig.~\ref{StatePreparation}(a)) appear between $10-11$ ms, demonstrating the state transfer.

The full state manipulation necessary for the trap loading is established with the addition of the 2nd OPL, as illustrated in Fig.~\ref{StatePreparation}(b).
The pink trace represents the LFS signal when the 2nd OPL pumps the HFS (between $9-10$ ms) at the saddle in addition to the 1st OPL laser. The replenishment of the LFS owing to the 2nd OPL is observed, yielding an efficiency of 15$\%$ for transferring LFS $\rightarrow$ HFS $\rightarrow$ LFS.

\subsection{Zeeman levels of CaF}

\begin{figure}[h!]
  \centering
    \includegraphics[width=3in]{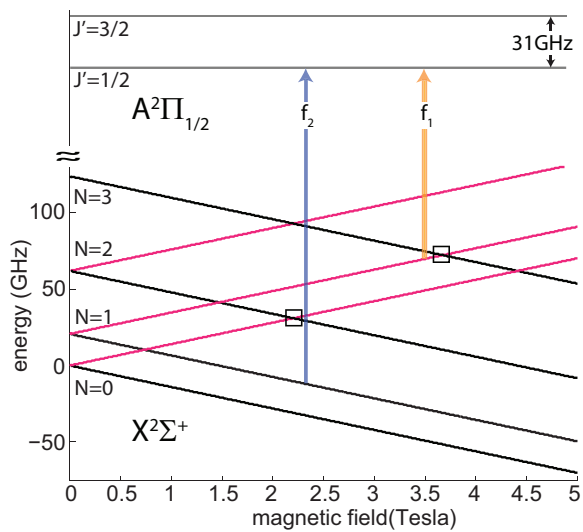}
    \caption{Zeeman levels of CaF in $X^2\Sigma^+$ ($v=0,N=0-3$) and $A^2\Pi_{1/2}$ ($v'=0,J'=1/2,3/2$). The low-field seeking and high-field seeking levels of $X^2\Sigma^+$ are plotted in pink and black and the Zeeman levels of $A^2\Pi_{1/2}$ are depicted in gray. Black squares indicate where the avoided crossings occur for the LFS of $N=0$ and $N=1$. The orange and blue arrows indicate the optical pumping fields for CaF ($N=1$) during trap loading.}
    \label{ZeemanLevel}
\end{figure}
We plot the Zeeman levels of CaF ($X^2\Sigma^+$) in pink (black) lines for LFS (HFS) up to 5 T in Fig.~\ref{ZeemanLevel}. Note that the low-field seeking state of the rotational level $N$ crosses the high-field seeking state of the $N+2$ state. The anisotropic hyperfine interaction causes these level crossings to become avoided crossings (inside the black squares in Fig.~\ref{ZeemanLevel}) at a field of $B_{\text{avoided}}=(E(N+2)-E(N))/2\mu_B$, where $E(N)$ is the rotational energy.
By scanning the frequency of the 1st OPL at different field strengths and measuring the LFS and HFS populations at the trap midplane, we map out the Zeeman levels of CaF and confirm the existence of the avoided crossing for $N=0$ at $B_{\text{avoided}}=2.2$ T.
When the CaF beam passes through $B_{\text{avoided}}$ on the way to the saddle, the LFS adiabatically turn into the HFS, inhibiting the operation of the trap at its full strength.
On the other hand, the LFS of $N=1$ have an increased $B_{\text{avoided}}$=3.67 T due to a larger rotational energy difference between $N=1$ and $N=3$, yielding a higher capture velocity than $N=0$. Therefore, we decide to perform the trapping experiment on $N=1$ at a field strength of 3.5 T.

Additionally, as pointed out by Ref.~\cite{Shuman2009}, the rotational leakage channel can be suppressed by driving a $N=1 \rightarrow J'=1/2$ transition (orange and blue arrows in Fig.~\ref{ZeemanLevel}), leading to a more efficient state transfer than $N=0$.
Since the 2nd OPL at frequency $f_2$ spatially overlaps with the CaF beam, it may excite the LFS to $A^2\Pi_{1/2} (v'=0, J'=3/2)$ at certain fields $B'$. This occurs when $f_2=f(N=1\rightarrow J'=1/2)+\mu_B B_2=f(N=1\rightarrow J'=3/2)-\mu_B B'$, where $f(N\rightarrow J')$ represents the transition frequencies at zero field.
Given the rotational splitting of the $A^2\Pi_{1/2}$ state, pumping the HFS at $B_2> 2.2$ T prevents the undesirable removal of the LFS.
To load CaF ($N=1$) into the trap, we choose ($B_1$, $B_2$)=(3.5, 2.27) T, yielding $v_c=29.8-33.5$ m/s and a trap depth of $E_D\sim$ 826 mK.

\subsection{Time decay signal with shutter incorporated}
\begin{figure}[h!]
  \centering
    \includegraphics[width=3in]{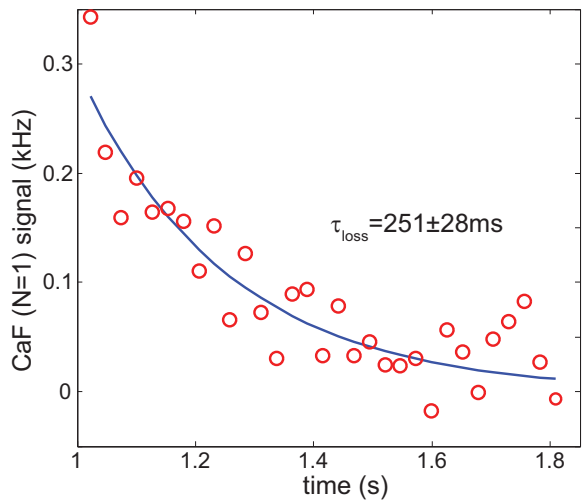}
    \caption{Time decay signal of CaF ($N=1$), $Sig(t')$, when the shutter is incorporated. The detection laser is switched on at $t_D=1$ s.}
    \label{WithShutterTimeDecayTrace}
\end{figure}
After the cryogenic shutter is closed, the trap loss of molecules results from the collisions with the background $^3$He gas. To observe this process, we use laser spectroscopy to monitor the trap decay of the molecules. Fig.~\ref{WithShutterTimeDecayTrace} shows the time decay signal of trapped CaF ($N=1$) when the shutter is incorporated, where the buffer-gas flow is 0.5 sccm during trap loading and the detection laser is switched on at $t_D=1$ s. The detection process itself is destructive due to decaying to the dark hyperfine states after spontaneous emissions and contributes a decay time constant of $\tau_{pumping}$ to the trapped sample. Therefore, the observed decay time constant of $\tau_{loss}=251\pm28$ ms can be expressed as $\tau_{loss}=(1/\tau_{pumping}+1/\tau_{BG})^{-1}$. In addition, the time decay signal can be written as $Sig(t')=Sig(0)\times e^{-t_D/\tau_{BG}}\times e^{-t'/\tau_{loss}}$, where $Sig(0)$ is the signal size immediately after the loading process and $t'$ is the delay time relative to $t_D$. By integrating the time decay signal over a duration of 200 ms (or $t'=0-200$ ms), we can extract $\tau_{BG}$ by plotting the integrated signal versus $t_D$, as shown in Fig.~3 of the main text.

\bibliography{OpticalLoadingPaperRef}

\begin{thebibliography}{44}
\expandafter\ifx\csname natexlab\endcsname\relax\def\natexlab#1{#1}\fi
\expandafter\ifx\csname bibnamefont\endcsname\relax
  \def\bibnamefont#1{#1}\fi
\expandafter\ifx\csname bibfnamefont\endcsname\relax
  \def\bibfnamefont#1{#1}\fi
\expandafter\ifx\csname citenamefont\endcsname\relax
  \def\citenamefont#1{#1}\fi
\expandafter\ifx\csname url\endcsname\relax
  \def\url#1{\texttt{#1}}\fi
\expandafter\ifx\csname urlprefix\endcsname\relax\def\urlprefix{URL }\fi
\providecommand{\bibinfo}[2]{#2}
\providecommand{\eprint}[2][]{\url{#2}}

\bibitem[{\citenamefont{Campbell et~al.}(2009)\citenamefont{Campbell,
  Tscherbul, Lu, Tsikata, Krems, and Doyle}}]{CampbellPRL2009}
\bibinfo{author}{\bibfnamefont{W.~C.} \bibnamefont{Campbell}},
  \bibinfo{author}{\bibfnamefont{T.~V.} \bibnamefont{Tscherbul}},
  \bibinfo{author}{\bibfnamefont{H.-I.} \bibnamefont{Lu}},
  \bibinfo{author}{\bibfnamefont{E.}~\bibnamefont{Tsikata}},
  \bibinfo{author}{\bibfnamefont{R.~V.} \bibnamefont{Krems}}, \bibnamefont{and}
  \bibinfo{author}{\bibfnamefont{J.~M.} \bibnamefont{Doyle}},
  \bibinfo{journal}{Phys. Rev. Lett.} \textbf{\bibinfo{volume}{102}},
  \bibinfo{pages}{013003} (\bibinfo{year}{2009}).

\bibitem[{\citenamefont{Hummon et~al.}(2011)\citenamefont{Hummon, Tscherbul,
  K{\l}os, Lu, Tsikata, Campbell, Dalgarno, and Doyle}}]{Hummon2011}
\bibinfo{author}{\bibfnamefont{M.~T.} \bibnamefont{Hummon}},
  \bibinfo{author}{\bibfnamefont{T.~V.} \bibnamefont{Tscherbul}},
  \bibinfo{author}{\bibfnamefont{J.}~\bibnamefont{K{\l}os}},
  \bibinfo{author}{\bibfnamefont{H.-I.} \bibnamefont{Lu}},
  \bibinfo{author}{\bibfnamefont{E.}~\bibnamefont{Tsikata}},
  \bibinfo{author}{\bibfnamefont{W.~C.} \bibnamefont{Campbell}},
  \bibinfo{author}{\bibfnamefont{A.}~\bibnamefont{Dalgarno}}, \bibnamefont{and}
  \bibinfo{author}{\bibfnamefont{J.~M.} \bibnamefont{Doyle}},
  \bibinfo{journal}{Phys. Rev. Lett.} \textbf{\bibinfo{volume}{106}},
  \bibinfo{pages}{053201} (\bibinfo{year}{2011}).

\bibitem[{\citenamefont{Zirbel et~al.}(2008)\citenamefont{Zirbel, Ni,
  Ospelkaus, D$\sp{\prime}$Incao, Wieman, Ye, and Jin}}]{Zirbel2008}
\bibinfo{author}{\bibfnamefont{J.~J.} \bibnamefont{Zirbel}},
  \bibinfo{author}{\bibfnamefont{K.-K.} \bibnamefont{Ni}},
  \bibinfo{author}{\bibfnamefont{S.}~\bibnamefont{Ospelkaus}},
  \bibinfo{author}{\bibfnamefont{J.~P.} \bibnamefont{D$\sp{\prime}$Incao}},
  \bibinfo{author}{\bibfnamefont{C.~E.} \bibnamefont{Wieman}},
  \bibinfo{author}{\bibfnamefont{J.}~\bibnamefont{Ye}}, \bibnamefont{and}
  \bibinfo{author}{\bibfnamefont{D.~S.} \bibnamefont{Jin}},
  \bibinfo{journal}{Phys. Rev. Lett.} \textbf{\bibinfo{volume}{100}},
  \bibinfo{pages}{143201} (\bibinfo{year}{2008}).

\bibitem[{\citenamefont{Ni et~al.}(2010)\citenamefont{Ni, Ospelkaus, Wang,
  Qu\'{e}m\'{e}ner, Neyenhuis, de~Miranda, Bohn, Ye, and Jin}}]{Ni2010}
\bibinfo{author}{\bibfnamefont{K.-K.} \bibnamefont{Ni}},
  \bibinfo{author}{\bibfnamefont{S.}~\bibnamefont{Ospelkaus}},
  \bibinfo{author}{\bibfnamefont{D.}~\bibnamefont{Wang}},
  \bibinfo{author}{\bibfnamefont{G.}~\bibnamefont{Qu\'{e}m\'{e}ner}},
  \bibinfo{author}{\bibfnamefont{B.}~\bibnamefont{Neyenhuis}},
  \bibinfo{author}{\bibfnamefont{M.~H.~G.} \bibnamefont{de~Miranda}},
  \bibinfo{author}{\bibfnamefont{J.~L.} \bibnamefont{Bohn}},
  \bibinfo{author}{\bibfnamefont{J.}~\bibnamefont{Ye}}, \bibnamefont{and}
  \bibinfo{author}{\bibfnamefont{D.~S.} \bibnamefont{Jin}},
  \bibinfo{journal}{Nature} \textbf{\bibinfo{volume}{464}},
  \bibinfo{pages}{1324} (\bibinfo{year}{2010}).

\bibitem[{\citenamefont{Sawyer et~al.}(2011)\citenamefont{Sawyer, Stuhl, Yeo,
  Tscherbul, Hummon, Xia, K{\l}os, Patterson, Doyle, and Ye}}]{Sawyer2011}
\bibinfo{author}{\bibfnamefont{B.~C.} \bibnamefont{Sawyer}},
  \bibinfo{author}{\bibfnamefont{B.~K.} \bibnamefont{Stuhl}},
  \bibinfo{author}{\bibfnamefont{M.}~\bibnamefont{Yeo}},
  \bibinfo{author}{\bibfnamefont{T.~V.} \bibnamefont{Tscherbul}},
  \bibinfo{author}{\bibfnamefont{M.~T.} \bibnamefont{Hummon}},
  \bibinfo{author}{\bibfnamefont{Y.}~\bibnamefont{Xia}},
  \bibinfo{author}{\bibfnamefont{J.}~\bibnamefont{K{\l}os}},
  \bibinfo{author}{\bibfnamefont{D.}~\bibnamefont{Patterson}},
  \bibinfo{author}{\bibfnamefont{J.~M.} \bibnamefont{Doyle}}, \bibnamefont{and}
  \bibinfo{author}{\bibfnamefont{J.}~\bibnamefont{Ye}}, \bibinfo{journal}{Phys.
  Chem. Chem. Phys.} \textbf{\bibinfo{volume}{13}}, \bibinfo{pages}{19059}
  (\bibinfo{year}{2011}).

\bibitem[{\citenamefont{Parazzoli et~al.}(2011)\citenamefont{Parazzoli, Fitch,
  \.{Z}uchowski, Hutson, and Lewandowski}}]{Parazzoli2011}
\bibinfo{author}{\bibfnamefont{L.~P.} \bibnamefont{Parazzoli}},
  \bibinfo{author}{\bibfnamefont{N.~J.} \bibnamefont{Fitch}},
  \bibinfo{author}{\bibfnamefont{P.~S.} \bibnamefont{\.{Z}uchowski}},
  \bibinfo{author}{\bibfnamefont{J.~M.} \bibnamefont{Hutson}},
  \bibnamefont{and} \bibinfo{author}{\bibfnamefont{H.~J.}
  \bibnamefont{Lewandowski}}, \bibinfo{journal}{Phys. Rev. Lett.}
  \textbf{\bibinfo{volume}{106}}, \bibinfo{pages}{193201}
  (\bibinfo{year}{2011}).

\bibitem[{\citenamefont{Stuhl et~al.}(2012)\citenamefont{Stuhl, Hummon, Yeo,
  Qu\'{e}m\'{e}ner, Bohn, and Ye}}]{Stuhl2012}
\bibinfo{author}{\bibfnamefont{B.~K.} \bibnamefont{Stuhl}},
  \bibinfo{author}{\bibfnamefont{M.~T.} \bibnamefont{Hummon}},
  \bibinfo{author}{\bibfnamefont{M.}~\bibnamefont{Yeo}},
  \bibinfo{author}{\bibfnamefont{G.}~\bibnamefont{Qu\'{e}m\'{e}ner}},
  \bibinfo{author}{\bibfnamefont{J.~L.} \bibnamefont{Bohn}}, \bibnamefont{and}
  \bibinfo{author}{\bibfnamefont{J.}~\bibnamefont{Ye}},
  \bibinfo{journal}{Nature} \textbf{\bibinfo{volume}{492}},
  \bibinfo{pages}{396} (\bibinfo{year}{2012}).

\bibitem[{\citenamefont{Kerman et~al.}(2004)\citenamefont{Kerman, Sage, Sainis,
  Bergeman, and DeMille}}]{Kerman2004}
\bibinfo{author}{\bibfnamefont{A.~J.} \bibnamefont{Kerman}},
  \bibinfo{author}{\bibfnamefont{J.~M.} \bibnamefont{Sage}},
  \bibinfo{author}{\bibfnamefont{S.}~\bibnamefont{Sainis}},
  \bibinfo{author}{\bibfnamefont{T.}~\bibnamefont{Bergeman}}, \bibnamefont{and}
  \bibinfo{author}{\bibfnamefont{D.}~\bibnamefont{DeMille}},
  \bibinfo{journal}{Phys. Rev. Lett.} \textbf{\bibinfo{volume}{92}},
  \bibinfo{pages}{033004} (\bibinfo{year}{2004}).

\bibitem[{\citenamefont{Ni et~al.}(2008)\citenamefont{Ni, Ospelkaus,
  de~Miranda, Pe$\sp{\prime}$er, Neyenhuis, Zirbel, Kotochigova, Julienne, Jin,
  and Ye}}]{Ni2008}
\bibinfo{author}{\bibfnamefont{K.-K.} \bibnamefont{Ni}},
  \bibinfo{author}{\bibfnamefont{S.}~\bibnamefont{Ospelkaus}},
  \bibinfo{author}{\bibfnamefont{M.~H.~G.} \bibnamefont{de~Miranda}},
  \bibinfo{author}{\bibfnamefont{A.}~\bibnamefont{Pe$\sp{\prime}$er}},
  \bibinfo{author}{\bibfnamefont{B.}~\bibnamefont{Neyenhuis}},
  \bibinfo{author}{\bibfnamefont{J.~J.} \bibnamefont{Zirbel}},
  \bibinfo{author}{\bibfnamefont{S.}~\bibnamefont{Kotochigova}},
  \bibinfo{author}{\bibfnamefont{P.~S.} \bibnamefont{Julienne}},
  \bibinfo{author}{\bibfnamefont{D.~S.} \bibnamefont{Jin}}, \bibnamefont{and}
  \bibinfo{author}{\bibfnamefont{J.}~\bibnamefont{Ye}},
  \bibinfo{journal}{Science} \textbf{\bibinfo{volume}{322}},
  \bibinfo{pages}{231} (\bibinfo{year}{2008}).

\bibitem[{\citenamefont{Ospelkaus et~al.}(2010)\citenamefont{Ospelkaus, Ni,
  Wang, de~Miranda, Neyenhuis, Qu{\'{e}}m{\'{e}}ner, Julienne, Bohn, Jin, and
  Ye}}]{Ospelkaus2010}
\bibinfo{author}{\bibfnamefont{S.}~\bibnamefont{Ospelkaus}},
  \bibinfo{author}{\bibfnamefont{K.-K.} \bibnamefont{Ni}},
  \bibinfo{author}{\bibfnamefont{D.}~\bibnamefont{Wang}},
  \bibinfo{author}{\bibfnamefont{M.~H.~G.} \bibnamefont{de~Miranda}},
  \bibinfo{author}{\bibfnamefont{B.}~\bibnamefont{Neyenhuis}},
  \bibinfo{author}{\bibfnamefont{G.}~\bibnamefont{Qu{\'{e}}m{\'{e}}ner}},
  \bibinfo{author}{\bibfnamefont{P.~S.} \bibnamefont{Julienne}},
  \bibinfo{author}{\bibfnamefont{J.~L.} \bibnamefont{Bohn}},
  \bibinfo{author}{\bibfnamefont{D.~S.} \bibnamefont{Jin}}, \bibnamefont{and}
  \bibinfo{author}{\bibfnamefont{J.}~\bibnamefont{Ye}},
  \bibinfo{journal}{Science} \textbf{\bibinfo{volume}{327}},
  \bibinfo{pages}{853} (\bibinfo{year}{2010}).

\bibitem[{\citenamefont{de~Miranda et~al.}(2011)\citenamefont{de~Miranda,
  Chotia, Neyenhuis, Wang, Qu\'{e}m\'{e}ner, Ospelkaus, Bohn, Ye, and
  Jin}}]{deMiranda2011}
\bibinfo{author}{\bibfnamefont{M.~H.~G.} \bibnamefont{de~Miranda}},
  \bibinfo{author}{\bibfnamefont{A.}~\bibnamefont{Chotia}},
  \bibinfo{author}{\bibfnamefont{B.}~\bibnamefont{Neyenhuis}},
  \bibinfo{author}{\bibfnamefont{D.}~\bibnamefont{Wang}},
  \bibinfo{author}{\bibfnamefont{G.}~\bibnamefont{Qu\'{e}m\'{e}ner}},
  \bibinfo{author}{\bibfnamefont{S.}~\bibnamefont{Ospelkaus}},
  \bibinfo{author}{\bibfnamefont{J.~L.} \bibnamefont{Bohn}},
  \bibinfo{author}{\bibfnamefont{J.}~\bibnamefont{Ye}}, \bibnamefont{and}
  \bibinfo{author}{\bibfnamefont{D.~S.} \bibnamefont{Jin}},
  \bibinfo{journal}{Nature Phys.} \textbf{\bibinfo{volume}{7}},
  \bibinfo{pages}{502} (\bibinfo{year}{2011}).

\bibitem[{\citenamefont{Vutha et~al.}(2010)\citenamefont{Vutha, Campbell,
  Gurevich, Hutzler, Parsons, Patterson, Petrik, Spaun, Doyle, Gabrielse
  et~al.}}]{Vutha2010}
\bibinfo{author}{\bibfnamefont{A.~C.} \bibnamefont{Vutha}},
  \bibinfo{author}{\bibfnamefont{W.~C.} \bibnamefont{Campbell}},
  \bibinfo{author}{\bibfnamefont{Y.~V.} \bibnamefont{Gurevich}},
  \bibinfo{author}{\bibfnamefont{N.~R.} \bibnamefont{Hutzler}},
  \bibinfo{author}{\bibfnamefont{M.}~\bibnamefont{Parsons}},
  \bibinfo{author}{\bibfnamefont{D.}~\bibnamefont{Patterson}},
  \bibinfo{author}{\bibfnamefont{E.}~\bibnamefont{Petrik}},
  \bibinfo{author}{\bibfnamefont{B.}~\bibnamefont{Spaun}},
  \bibinfo{author}{\bibfnamefont{J.~M.} \bibnamefont{Doyle}},
  \bibinfo{author}{\bibfnamefont{G.}~\bibnamefont{Gabrielse}},
  \bibnamefont{et~al.}, \bibinfo{journal}{J. Phys. B}
  \textbf{\bibinfo{volume}{43}}, \bibinfo{pages}{074007}
  (\bibinfo{year}{2010}).

\bibitem[{\citenamefont{Hudson et~al.}(2011)\citenamefont{Hudson, Kara,
  Smallman, Sauer, Tarbutt, and Hinds}}]{Hudson2011}
\bibinfo{author}{\bibfnamefont{J.~J.} \bibnamefont{Hudson}},
  \bibinfo{author}{\bibfnamefont{D.~M.} \bibnamefont{Kara}},
  \bibinfo{author}{\bibfnamefont{I.~J.} \bibnamefont{Smallman}},
  \bibinfo{author}{\bibfnamefont{B.~E.} \bibnamefont{Sauer}},
  \bibinfo{author}{\bibfnamefont{M.~R.} \bibnamefont{Tarbutt}},
  \bibnamefont{and} \bibinfo{author}{\bibfnamefont{E.~A.} \bibnamefont{Hinds}},
  \bibinfo{journal}{Nature} \textbf{\bibinfo{volume}{473}},
  \bibinfo{pages}{493} (\bibinfo{year}{2011}).

\bibitem[{\citenamefont{Lemeshko et~al.}(2013)\citenamefont{Lemeshko, Krems,
  Doyle, and Kais}}]{Lemeshko2013}
\bibinfo{author}{\bibfnamefont{M.}~\bibnamefont{Lemeshko}},
  \bibinfo{author}{\bibfnamefont{R.~V.} \bibnamefont{Krems}},
  \bibinfo{author}{\bibfnamefont{J.~M.} \bibnamefont{Doyle}}, \bibnamefont{and}
  \bibinfo{author}{\bibfnamefont{S.}~\bibnamefont{Kais}},
  \bibinfo{journal}{Molecular Physics} \textbf{\bibinfo{volume}{111}},
  \bibinfo{pages}{1648} (\bibinfo{year}{2013}).

\bibitem[{\citenamefont{van~de Meerakker et~al.}(2005)\citenamefont{van~de
  Meerakker, Smeets, Vanhaecke, Jongma, and Meijer}}]{Meerakker2005}
\bibinfo{author}{\bibfnamefont{S.~Y.~T.} \bibnamefont{van~de Meerakker}},
  \bibinfo{author}{\bibfnamefont{P.~H.~M.} \bibnamefont{Smeets}},
  \bibinfo{author}{\bibfnamefont{N.}~\bibnamefont{Vanhaecke}},
  \bibinfo{author}{\bibfnamefont{R.~T.} \bibnamefont{Jongma}},
  \bibnamefont{and} \bibinfo{author}{\bibfnamefont{G.}~\bibnamefont{Meijer}},
  \bibinfo{journal}{Phys. Rev. Lett.} \textbf{\bibinfo{volume}{94}},
  \bibinfo{pages}{023004} (\bibinfo{year}{2005}).

\bibitem[{\citenamefont{Hoekstra et~al.}(2007)\citenamefont{Hoekstra,
  Mets\"{a}l\"{a}, Zieger, Scharfenberg, Gilijamse, Meijer, and van~de
  Meerakker}}]{Hoekstra2007}
\bibinfo{author}{\bibfnamefont{S.}~\bibnamefont{Hoekstra}},
  \bibinfo{author}{\bibfnamefont{M.}~\bibnamefont{Mets\"{a}l\"{a}}},
  \bibinfo{author}{\bibfnamefont{P.~C.} \bibnamefont{Zieger}},
  \bibinfo{author}{\bibfnamefont{L.}~\bibnamefont{Scharfenberg}},
  \bibinfo{author}{\bibfnamefont{J.~J.} \bibnamefont{Gilijamse}},
  \bibinfo{author}{\bibfnamefont{G.}~\bibnamefont{Meijer}}, \bibnamefont{and}
  \bibinfo{author}{\bibfnamefont{S.~Y.~T.} \bibnamefont{van~de Meerakker}},
  \bibinfo{journal}{Phys. Rev. A} \textbf{\bibinfo{volume}{76}},
  \bibinfo{pages}{063408} (\bibinfo{year}{2007}).

\bibitem[{\citenamefont{Momose et~al.}(2013)\citenamefont{Momose, Liu, Zhou,
  Djuricanin, and Carty}}]{Momose2013}
\bibinfo{author}{\bibfnamefont{T.}~\bibnamefont{Momose}},
  \bibinfo{author}{\bibfnamefont{Y.}~\bibnamefont{Liu}},
  \bibinfo{author}{\bibfnamefont{S.}~\bibnamefont{Zhou}},
  \bibinfo{author}{\bibfnamefont{P.}~\bibnamefont{Djuricanin}},
  \bibnamefont{and} \bibinfo{author}{\bibfnamefont{D.}~\bibnamefont{Carty}},
  \bibinfo{journal}{Phys. Chem. Chem. Phys.} \textbf{\bibinfo{volume}{15}},
  \bibinfo{pages}{1772} (\bibinfo{year}{2013}).

\bibitem[{\citenamefont{Narevicius et~al.}(2008)\citenamefont{Narevicius,
  Libson, Parthey, Chavez, Narevicius, Even, and Raizen}}]{Narevicius2008_II}
\bibinfo{author}{\bibfnamefont{E.}~\bibnamefont{Narevicius}},
  \bibinfo{author}{\bibfnamefont{A.}~\bibnamefont{Libson}},
  \bibinfo{author}{\bibfnamefont{C.~G.} \bibnamefont{Parthey}},
  \bibinfo{author}{\bibfnamefont{I.}~\bibnamefont{Chavez}},
  \bibinfo{author}{\bibfnamefont{J.}~\bibnamefont{Narevicius}},
  \bibinfo{author}{\bibfnamefont{U.}~\bibnamefont{Even}}, \bibnamefont{and}
  \bibinfo{author}{\bibfnamefont{M.~G.} \bibnamefont{Raizen}},
  \bibinfo{journal}{Phys. Rev. A} \textbf{\bibinfo{volume}{77}},
  \bibinfo{pages}{051401} (\bibinfo{year}{2008}).

\bibitem[{\citenamefont{van Buuren et~al.}(2009)\citenamefont{van Buuren,
  Sommer, Motsch, Pohle, Schenk, Bayerl, Pinkse, and Rempe}}]{vanBuuren2009}
\bibinfo{author}{\bibfnamefont{L.~D.} \bibnamefont{van Buuren}},
  \bibinfo{author}{\bibfnamefont{C.}~\bibnamefont{Sommer}},
  \bibinfo{author}{\bibfnamefont{M.}~\bibnamefont{Motsch}},
  \bibinfo{author}{\bibfnamefont{S.}~\bibnamefont{Pohle}},
  \bibinfo{author}{\bibfnamefont{M.}~\bibnamefont{Schenk}},
  \bibinfo{author}{\bibfnamefont{J.}~\bibnamefont{Bayerl}},
  \bibinfo{author}{\bibfnamefont{P.~W.~H.} \bibnamefont{Pinkse}},
  \bibnamefont{and} \bibinfo{author}{\bibfnamefont{G.}~\bibnamefont{Rempe}},
  \bibinfo{journal}{Phys. Rev. Lett.} \textbf{\bibinfo{volume}{102}},
  \bibinfo{pages}{033001} (\bibinfo{year}{2009}).

\bibitem[{\citenamefont{Chervenkov et~al.}(2014)\citenamefont{Chervenkov, Wu,
  Bayerl, Rohlfes, Gantner, Zeppenfeld, and Rempe}}]{Chervenkov2014}
\bibinfo{author}{\bibfnamefont{S.}~\bibnamefont{Chervenkov}},
  \bibinfo{author}{\bibfnamefont{X.}~\bibnamefont{Wu}},
  \bibinfo{author}{\bibfnamefont{J.}~\bibnamefont{Bayerl}},
  \bibinfo{author}{\bibfnamefont{A.}~\bibnamefont{Rohlfes}},
  \bibinfo{author}{\bibfnamefont{T.}~\bibnamefont{Gantner}},
  \bibinfo{author}{\bibfnamefont{M.}~\bibnamefont{Zeppenfeld}},
  \bibnamefont{and} \bibinfo{author}{\bibfnamefont{G.}~\bibnamefont{Rempe}},
  \bibinfo{journal}{Phys. Rev. Lett.} \textbf{\bibinfo{volume}{112}},
  \bibinfo{pages}{013001} (\bibinfo{year}{2014}).

\bibitem[{\citenamefont{Hutzler et~al.}(2011)\citenamefont{Hutzler, Parsons,
  Gurevich, Hess, Petrik, Spaun, Vutha, DeMille, Gabrielse, and
  Doyle}}]{Hutzler2011}
\bibinfo{author}{\bibfnamefont{N.~R.} \bibnamefont{Hutzler}},
  \bibinfo{author}{\bibfnamefont{M.~F.} \bibnamefont{Parsons}},
  \bibinfo{author}{\bibfnamefont{Y.~V.} \bibnamefont{Gurevich}},
  \bibinfo{author}{\bibfnamefont{P.~W.} \bibnamefont{Hess}},
  \bibinfo{author}{\bibfnamefont{E.}~\bibnamefont{Petrik}},
  \bibinfo{author}{\bibfnamefont{B.}~\bibnamefont{Spaun}},
  \bibinfo{author}{\bibfnamefont{A.~C.} \bibnamefont{Vutha}},
  \bibinfo{author}{\bibfnamefont{D.}~\bibnamefont{DeMille}},
  \bibinfo{author}{\bibfnamefont{G.}~\bibnamefont{Gabrielse}},
  \bibnamefont{and} \bibinfo{author}{\bibfnamefont{J.~M.} \bibnamefont{Doyle}},
  \bibinfo{journal}{Phys. Chem. Chem. Phys.} \textbf{\bibinfo{volume}{13}},
  \bibinfo{pages}{18976} (\bibinfo{year}{2011}).

\bibitem[{\citenamefont{Barry et~al.}(2011)\citenamefont{Barry, Shuman, and
  DeMille}}]{Barry2011}
\bibinfo{author}{\bibfnamefont{J.~F.} \bibnamefont{Barry}},
  \bibinfo{author}{\bibfnamefont{E.~S.} \bibnamefont{Shuman}},
  \bibnamefont{and} \bibinfo{author}{\bibfnamefont{D.}~\bibnamefont{DeMille}},
  \bibinfo{journal}{Phys. Chem. Chem. Phys.} \textbf{\bibinfo{volume}{13}},
  \bibinfo{pages}{18936} (\bibinfo{year}{2011}).

\bibitem[{\citenamefont{Bethlem et~al.}(2002)\citenamefont{Bethlem, Crompvoets,
  Jongma, van~de Meerakker, and Meijer}}]{Bethlem2002}
\bibinfo{author}{\bibfnamefont{H.~L.} \bibnamefont{Bethlem}},
  \bibinfo{author}{\bibfnamefont{F.~M.~H.} \bibnamefont{Crompvoets}},
  \bibinfo{author}{\bibfnamefont{R.~T.} \bibnamefont{Jongma}},
  \bibinfo{author}{\bibfnamefont{S.~Y.~T.} \bibnamefont{van~de Meerakker}},
  \bibnamefont{and} \bibinfo{author}{\bibfnamefont{G.}~\bibnamefont{Meijer}},
  \bibinfo{journal}{Phys. Rev. A} \textbf{\bibinfo{volume}{65}},
  \bibinfo{pages}{053416} (\bibinfo{year}{2002}).

\bibitem[{\citenamefont{Englert et~al.}(2011)\citenamefont{Englert, Mielenz,
  Sommer, Bayerl, Motsch, Pinkse, Rempe, and Zeppenfeld}}]{Englert2011}
\bibinfo{author}{\bibfnamefont{B.~G.~U.} \bibnamefont{Englert}},
  \bibinfo{author}{\bibfnamefont{M.}~\bibnamefont{Mielenz}},
  \bibinfo{author}{\bibfnamefont{C.}~\bibnamefont{Sommer}},
  \bibinfo{author}{\bibfnamefont{J.}~\bibnamefont{Bayerl}},
  \bibinfo{author}{\bibfnamefont{M.}~\bibnamefont{Motsch}},
  \bibinfo{author}{\bibfnamefont{P.~W.~H.} \bibnamefont{Pinkse}},
  \bibinfo{author}{\bibfnamefont{G.}~\bibnamefont{Rempe}}, \bibnamefont{and}
  \bibinfo{author}{\bibfnamefont{M.}~\bibnamefont{Zeppenfeld}},
  \bibinfo{journal}{Phys. Rev. Lett.} \textbf{\bibinfo{volume}{107}},
  \bibinfo{pages}{263003} (\bibinfo{year}{2011}).

\bibitem[{\citenamefont{Zeppenfeld et~al.}(2012)\citenamefont{Zeppenfeld,
  Englert, Gl\"{o}ckner, Prehn, Mielenz, Sommer, van Buuren, Motsch, and
  Rempe}}]{Zeppenfeld2012}
\bibinfo{author}{\bibfnamefont{M.}~\bibnamefont{Zeppenfeld}},
  \bibinfo{author}{\bibfnamefont{B.~G.~U.} \bibnamefont{Englert}},
  \bibinfo{author}{\bibfnamefont{R.}~\bibnamefont{Gl\"{o}ckner}},
  \bibinfo{author}{\bibfnamefont{A.}~\bibnamefont{Prehn}},
  \bibinfo{author}{\bibfnamefont{M.}~\bibnamefont{Mielenz}},
  \bibinfo{author}{\bibfnamefont{C.}~\bibnamefont{Sommer}},
  \bibinfo{author}{\bibfnamefont{L.~D.} \bibnamefont{van Buuren}},
  \bibinfo{author}{\bibfnamefont{M.}~\bibnamefont{Motsch}}, \bibnamefont{and}
  \bibinfo{author}{\bibfnamefont{G.}~\bibnamefont{Rempe}},
  \bibinfo{journal}{Nature} \textbf{\bibinfo{volume}{491}},
  \bibinfo{pages}{570} (\bibinfo{year}{2012}).

\bibitem[{\citenamefont{Hutzler et~al.}(2012)\citenamefont{Hutzler, Lu, and
  Doyle}}]{Hutzler2012}
\bibinfo{author}{\bibfnamefont{N.~R.} \bibnamefont{Hutzler}},
  \bibinfo{author}{\bibfnamefont{H.-I.} \bibnamefont{Lu}}, \bibnamefont{and}
  \bibinfo{author}{\bibfnamefont{J.~M.} \bibnamefont{Doyle}},
  \bibinfo{journal}{Chem. Rev.} \textbf{\bibinfo{volume}{112}},
  \bibinfo{pages}{4803} (\bibinfo{year}{2012}).

\bibitem[{\citenamefont{Cornell et~al.}(1991)\citenamefont{Cornell, Monroe, and
  Wieman}}]{Cornell1991}
\bibinfo{author}{\bibfnamefont{E.~A.} \bibnamefont{Cornell}},
  \bibinfo{author}{\bibfnamefont{C.}~\bibnamefont{Monroe}}, \bibnamefont{and}
  \bibinfo{author}{\bibfnamefont{C.~E.} \bibnamefont{Wieman}},
  \bibinfo{journal}{Phys. Rev. Lett.} \textbf{\bibinfo{volume}{67}},
  \bibinfo{pages}{2439} (\bibinfo{year}{1991}).

\bibitem[{\citenamefont{Falkenau et~al.}(2011)\citenamefont{Falkenau, Volchkov,
  R\"{u}hrig, Griesmaier, and Pfau}}]{Falkenau2011}
\bibinfo{author}{\bibfnamefont{M.}~\bibnamefont{Falkenau}},
  \bibinfo{author}{\bibfnamefont{V.~V.} \bibnamefont{Volchkov}},
  \bibinfo{author}{\bibfnamefont{J.}~\bibnamefont{R\"{u}hrig}},
  \bibinfo{author}{\bibfnamefont{A.}~\bibnamefont{Griesmaier}},
  \bibnamefont{and} \bibinfo{author}{\bibfnamefont{T.}~\bibnamefont{Pfau}},
  \bibinfo{journal}{Phys. Rev. Lett.} \textbf{\bibinfo{volume}{106}},
  \bibinfo{pages}{163002} (\bibinfo{year}{2011}).

\bibitem[{\citenamefont{Campbell and Doyle}(2009)}]{Campbell2009}
\bibinfo{author}{\bibfnamefont{W.~C.} \bibnamefont{Campbell}} \bibnamefont{and}
  \bibinfo{author}{\bibfnamefont{J.~M.} \bibnamefont{Doyle}},
  \emph{\bibinfo{title}{{Cold molecules: Theory, experiment, applications}}}
  (\bibinfo{publisher}{CRC Press}, \bibinfo{year}{2009}),
  chap.~\bibinfo{chapter}{13}.

\bibitem[{\citenamefont{Weinstein et~al.}(1998)\citenamefont{Weinstein,
  deCarvalho, Guillet, Friedrich, and Doyle}}]{Weinstein1998}
\bibinfo{author}{\bibfnamefont{J.~D.} \bibnamefont{Weinstein}},
  \bibinfo{author}{\bibfnamefont{R.}~\bibnamefont{deCarvalho}},
  \bibinfo{author}{\bibfnamefont{T.}~\bibnamefont{Guillet}},
  \bibinfo{author}{\bibfnamefont{B.}~\bibnamefont{Friedrich}},
  \bibnamefont{and} \bibinfo{author}{\bibfnamefont{J.~M.} \bibnamefont{Doyle}},
  \bibinfo{journal}{Nature} \textbf{\bibinfo{volume}{395}},
  \bibinfo{pages}{148} (\bibinfo{year}{1998}).

\bibitem[{\citenamefont{Campbell et~al.}(2007)\citenamefont{Campbell, Tsikata,
  Lu, van Buuren, and Doyle}}]{Campbell2007}
\bibinfo{author}{\bibfnamefont{W.~C.} \bibnamefont{Campbell}},
  \bibinfo{author}{\bibfnamefont{E.}~\bibnamefont{Tsikata}},
  \bibinfo{author}{\bibfnamefont{H.-I.} \bibnamefont{Lu}},
  \bibinfo{author}{\bibfnamefont{L.~D.} \bibnamefont{van Buuren}},
  \bibnamefont{and} \bibinfo{author}{\bibfnamefont{J.~M.} \bibnamefont{Doyle}},
  \bibinfo{journal}{Phys. Rev. Lett.} \textbf{\bibinfo{volume}{98}},
  \bibinfo{pages}{213001} (\bibinfo{year}{2007}).

\bibitem[{\citenamefont{Stoll et~al.}(2008)\citenamefont{Stoll, Bakker,
  Steimle, Meijer, and Peters}}]{Stoll2008}
\bibinfo{author}{\bibfnamefont{M.}~\bibnamefont{Stoll}},
  \bibinfo{author}{\bibfnamefont{J.~M.} \bibnamefont{Bakker}},
  \bibinfo{author}{\bibfnamefont{T.~C.} \bibnamefont{Steimle}},
  \bibinfo{author}{\bibfnamefont{G.}~\bibnamefont{Meijer}}, \bibnamefont{and}
  \bibinfo{author}{\bibfnamefont{A.}~\bibnamefont{Peters}},
  \bibinfo{journal}{Phys. Rev. A} \textbf{\bibinfo{volume}{78}},
  \bibinfo{pages}{032707} (\bibinfo{year}{2008}).

\bibitem[{\citenamefont{Tsikata et~al.}(2010)\citenamefont{Tsikata, Campbell,
  Hummon, Lu, and Doyle}}]{Tsikata2010}
\bibinfo{author}{\bibfnamefont{E.}~\bibnamefont{Tsikata}},
  \bibinfo{author}{\bibfnamefont{W.~C.} \bibnamefont{Campbell}},
  \bibinfo{author}{\bibfnamefont{M.~T.} \bibnamefont{Hummon}},
  \bibinfo{author}{\bibfnamefont{H.-I.} \bibnamefont{Lu}}, \bibnamefont{and}
  \bibinfo{author}{\bibfnamefont{J.~M.} \bibnamefont{Doyle}},
  \bibinfo{journal}{New J. Phys.} \textbf{\bibinfo{volume}{12}},
  \bibinfo{pages}{065028} (\bibinfo{year}{2010}).

\bibitem[{\citenamefont{Hancox et~al.}(2004)\citenamefont{Hancox, Doret,
  Hummon, Luo, and Doyle}}]{Hancox2004}
\bibinfo{author}{\bibfnamefont{C.~I.} \bibnamefont{Hancox}},
  \bibinfo{author}{\bibfnamefont{S.~C.} \bibnamefont{Doret}},
  \bibinfo{author}{\bibfnamefont{M.~T.} \bibnamefont{Hummon}},
  \bibinfo{author}{\bibfnamefont{L.}~\bibnamefont{Luo}}, \bibnamefont{and}
  \bibinfo{author}{\bibfnamefont{J.~M.} \bibnamefont{Doyle}},
  \bibinfo{journal}{Nature} \textbf{\bibinfo{volume}{431}},
  \bibinfo{pages}{281} (\bibinfo{year}{2004}).

\bibitem[{\citenamefont{Shuman et~al.}(2010)\citenamefont{Shuman, Barry, and
  Demille}}]{Shuman2010}
\bibinfo{author}{\bibfnamefont{E.~S.} \bibnamefont{Shuman}},
  \bibinfo{author}{\bibfnamefont{J.~F.} \bibnamefont{Barry}}, \bibnamefont{and}
  \bibinfo{author}{\bibfnamefont{D.}~\bibnamefont{Demille}},
  \bibinfo{journal}{Nature} \textbf{\bibinfo{volume}{467}},
  \bibinfo{pages}{820} (\bibinfo{year}{2010}).

\bibitem[{\citenamefont{Barry et~al.}(2012)\citenamefont{Barry, Shuman,
  Norrgard, and Demille}}]{Barry2012}
\bibinfo{author}{\bibfnamefont{J.~F.} \bibnamefont{Barry}},
  \bibinfo{author}{\bibfnamefont{E.~S.} \bibnamefont{Shuman}},
  \bibinfo{author}{\bibfnamefont{E.~B.} \bibnamefont{Norrgard}},
  \bibnamefont{and} \bibinfo{author}{\bibfnamefont{D.}~\bibnamefont{Demille}},
  \bibinfo{journal}{Phys. Rev. Lett.} \textbf{\bibinfo{volume}{108}},
  \bibinfo{pages}{103002} (\bibinfo{year}{2012}).

\bibitem[{\citenamefont{Hummon et~al.}(2013)\citenamefont{Hummon, Yeo, Stuhl,
  Collopy, Xia, and Ye}}]{Hummon2013}
\bibinfo{author}{\bibfnamefont{M.~T.} \bibnamefont{Hummon}},
  \bibinfo{author}{\bibfnamefont{M.}~\bibnamefont{Yeo}},
  \bibinfo{author}{\bibfnamefont{B.~K.} \bibnamefont{Stuhl}},
  \bibinfo{author}{\bibfnamefont{A.~L.} \bibnamefont{Collopy}},
  \bibinfo{author}{\bibfnamefont{Y.}~\bibnamefont{Xia}}, \bibnamefont{and}
  \bibinfo{author}{\bibfnamefont{J.}~\bibnamefont{Ye}}, \bibinfo{journal}{Phys.
  Rev. Lett.} \textbf{\bibinfo{volume}{110}}, \bibinfo{pages}{143001}
  (\bibinfo{year}{2013}).

\bibitem[{\citenamefont{Lu et~al.}(2011)\citenamefont{Lu, Rasmussen, Wright,
  Patterson, and Doyle}}]{Lu2011}
\bibinfo{author}{\bibfnamefont{H.-I.} \bibnamefont{Lu}},
  \bibinfo{author}{\bibfnamefont{J.}~\bibnamefont{Rasmussen}},
  \bibinfo{author}{\bibfnamefont{M.~J.} \bibnamefont{Wright}},
  \bibinfo{author}{\bibfnamefont{D.}~\bibnamefont{Patterson}},
  \bibnamefont{and} \bibinfo{author}{\bibfnamefont{J.~M.} \bibnamefont{Doyle}},
  \bibinfo{journal}{Phys. Chem. Chem. Phys.} \textbf{\bibinfo{volume}{13}},
  \bibinfo{pages}{18986} (\bibinfo{year}{2011}).

\bibitem[{SM2()}]{SM2013}
\bibinfo{note}{See supplemental material provided.}

\bibitem[{\citenamefont{Shuman et~al.}(2009)\citenamefont{Shuman, Barry, Glenn,
  and DeMille}}]{Shuman2009}
\bibinfo{author}{\bibfnamefont{E.~S.} \bibnamefont{Shuman}},
  \bibinfo{author}{\bibfnamefont{J.~F.} \bibnamefont{Barry}},
  \bibinfo{author}{\bibfnamefont{D.~R.} \bibnamefont{Glenn}}, \bibnamefont{and}
  \bibinfo{author}{\bibfnamefont{D.}~\bibnamefont{DeMille}},
  \bibinfo{journal}{Phys. Rev. Lett.} \textbf{\bibinfo{volume}{103}},
  \bibinfo{pages}{223001} (\bibinfo{year}{2009}).

\bibitem[{\citenamefont{Maussang et~al.}(2005)\citenamefont{Maussang, Egorov,
  Helton, Nguyen, and Doyle}}]{Maussang2005}
\bibinfo{author}{\bibfnamefont{K.}~\bibnamefont{Maussang}},
  \bibinfo{author}{\bibfnamefont{D.}~\bibnamefont{Egorov}},
  \bibinfo{author}{\bibfnamefont{J.~S.} \bibnamefont{Helton}},
  \bibinfo{author}{\bibfnamefont{S.~V.} \bibnamefont{Nguyen}},
  \bibnamefont{and} \bibinfo{author}{\bibfnamefont{J.~M.} \bibnamefont{Doyle}},
  \bibinfo{journal}{Phys. Rev. Lett.} \textbf{\bibinfo{volume}{94}},
  \bibinfo{pages}{123002} (\bibinfo{year}{2005}).

\bibitem[{\citenamefont{Tscherbul
  et~al.}(2011{\natexlab{a}})\citenamefont{Tscherbul, K{\l}os, and
  Buchachenko}}]{Tscherbul2011}
\bibinfo{author}{\bibfnamefont{T.~V.} \bibnamefont{Tscherbul}},
  \bibinfo{author}{\bibfnamefont{J.}~\bibnamefont{K{\l}os}}, \bibnamefont{and}
  \bibinfo{author}{\bibfnamefont{A.~A.} \bibnamefont{Buchachenko}},
  \bibinfo{journal}{Phys. Rev. A} \textbf{\bibinfo{volume}{84}},
  \bibinfo{pages}{040701} (\bibinfo{year}{2011}{\natexlab{a}}).

\bibitem[{\citenamefont{Singh et~al.}(2012)\citenamefont{Singh, Hardman, Tariq,
  Lu, Ellis, Morrison, and Weinstein}}]{Singh2012}
\bibinfo{author}{\bibfnamefont{V.}~\bibnamefont{Singh}},
  \bibinfo{author}{\bibfnamefont{K.~S.} \bibnamefont{Hardman}},
  \bibinfo{author}{\bibfnamefont{N.}~\bibnamefont{Tariq}},
  \bibinfo{author}{\bibfnamefont{M.-J.} \bibnamefont{Lu}},
  \bibinfo{author}{\bibfnamefont{A.}~\bibnamefont{Ellis}},
  \bibinfo{author}{\bibfnamefont{M.~J.} \bibnamefont{Morrison}},
  \bibnamefont{and} \bibinfo{author}{\bibfnamefont{J.~D.}
  \bibnamefont{Weinstein}}, \bibinfo{journal}{Phys. Rev. Lett.}
  \textbf{\bibinfo{volume}{108}}, \bibinfo{pages}{203201}
  (\bibinfo{year}{2012}).

\bibitem[{\citenamefont{Tscherbul
  et~al.}(2011{\natexlab{b}})\citenamefont{Tscherbul, Yu, and
  Dalgarno}}]{Tscherbul2011sympathetic}
\bibinfo{author}{\bibfnamefont{T.~V.} \bibnamefont{Tscherbul}},
  \bibinfo{author}{\bibfnamefont{H.-G.} \bibnamefont{Yu}}, \bibnamefont{and}
  \bibinfo{author}{\bibfnamefont{A.}~\bibnamefont{Dalgarno}},
  \bibinfo{journal}{Phys. Rev. Lett.} \textbf{\bibinfo{volume}{106}},
  \bibinfo{pages}{073201} (\bibinfo{year}{2011}{\natexlab{b}}).

\end{thebibliography}

\end{document}